\documentclass[12pt,oneside, a4paper]{article}

\ifx\pdfoutput\undefined
\usepackage[dvips,bookmarks=false]{hyperref}	
\else
\usepackage{hyperref}	
\fi
\hypersetup{colorlinks,bookmarksopen,bookmarksnumbered,citecolor=blue,
linkcolor=blue,pdfstartview=FitH,urlcolor=blue}

\usepackage{graphicx}
\usepackage{amssymb}
\usepackage{cite}
\usepackage{bm}
\usepackage{amsmath,amsthm}

\newcommand{\capdef}{}
\newcommand{\mycaption}[2][\capdef]{\renewcommand{\capdef}{#2}%
       \caption[#1]{{\footnotesize #2}}}

\newcommand{\be}{\begin{equation}}
\newcommand{\ee}{\end{equation}}

\newcommand{\dmq}{\ensuremath{\Delta m^2_{31}}}
\newcommand{\sq}{\ensuremath{\sin^22\theta_{13}}}

\begin{document}

\begin{titlepage}

\begin{center}

\vspace{1cm}
{\Large\bf Identifying the Neutrino mass Ordering\\[2mm] 
with INO and NOvA}
\vspace{1cm}

\renewcommand{\thefootnote}{\fnsymbol{footnote}}
{\bf Mattias Blennow}\footnote[1]{blennow AT mpi-hd.mpg.de}, 
{\bf Thomas Schwetz}\footnote[2]{schwetz AT mpi-hd.mpg.de},
\vspace{5mm}

{\it%
{Max-Planck-Institut f\"ur Kernphysik, Saupfercheckweg 1, 69117 Heidelberg, Germany}}

\vspace{8mm} 

\abstract{The relatively large value of $\theta_{13}$ established recently
by the Daya Bay reactor experiment opens the possibility to determine the
neutrino mass ordering with experiments currently under construction. We
investigate synergies between the NOvA long-baseline accelerator experiment
with atmospheric neutrino data from the India-based Neutrino Observatory
(INO). We identify the requirements on energy and direction reconstruction
and detector mass for INO necessary for a significant sensitivity. If
neutrino energy and direction reconstruction at the level of 10\% and
$10^\circ$ can be achieved by INO a determination of the neutrino mass
ordering seems possible around 2020.}

\end{center}
\end{titlepage}

\renewcommand{\thefootnote}{\arabic{footnote}}
\setcounter{footnote}{0}

\setcounter{page}{2}

\section{Introduction}

Huge progress has been achieved in the study of neutrino
oscillations~\cite{Fukuda:1998mi, Ahmad:2002jz, Apollonio:2002gd, Araki:2004mb,
  Adamson:2008zt} and a rough picture of the structure of
three-flavour lepton mixing has been obtained, with two large mixing
angles ($\theta_{12}$ and $\theta_{23}$) and two neutrino mass-squared
differences separated roughly by a factor 30: $\Delta m^2_{21} \simeq
7.6\cdot 10^{-5}$~eV$^2$ and $|\Delta m^2_{31}| \simeq 2.4\cdot
10^{-3}$~eV$^2$, see \cite{Schwetz:2011qt, Schwetz:2011zk} for a
recent global fit. The sign of $\Delta m^2_{21}$ is determined by
the matter effect \cite{Wolfenstein:1977ue, Barger:1980tf,
  Mikheev:1986gs} inside the sun being responsible for the flavour
transition of solar neutrinos.\footnote{The convention-independent
  statement is that the neutrino mass state which contains dominantly
  the electron neutrino (denoted by $\nu_1$ per convention) has to be
  the lighter of the two mass states responsible for the flavour
  transitions observed for solar neutrinos and reactor anti-neutrinos in
  the KamLAND experiment.} In contrast the sign of $\Delta m^2_{31}$
is not known and present data cannot distinguish between the so-called
normal or inverted neutrino mass ordering, with $\Delta m^2_{31} > 0$
or $\Delta m^2_{31} < 0$, respectively. The determination of the sign
of $\Delta m^2_{31}$ is one of the most important goals of the future
neutrino oscillation program. The type of the neutrino mass ordering
provides crucial information on the flavour structure in the lepton
sector. Furthermore, the unknown sign of \dmq\ introduces a two-fold
ambiguity in the determination of the parameters $\theta_{13}$ and
$\delta$ by long-baseline experiments~\cite{Minakata:2001qm, Barger:2001yr},
and might severely affect the search for CP violation in neutrino
oscillations.

The most promising way to determine the neutrino mass ordering is to search
for matter effects \cite{Wolfenstein:1977ue, Barger:1980tf, Mikheev:1986gs}
in oscillations driven by \dmq. This requires the participation of the
electron neutrino in such oscillations, which is suppressed by the mixing
angle $\theta_{13}$. Therefore, the size of $\theta_{13}$ strongly affects
the possibility to determine the sign of \dmq.  Recently the Daya Bay
reactor neutrino experiment established a non-zero value of $\theta_{13}$ at
the $5\sigma$ level~\cite{dayabay}, 
\begin{equation}\label{eq:th13}
\sq = 0.092 \pm 0.016({\rm stat}) \pm 0.005 ({\rm syst}) \,. 
\end{equation}
A similar result was later found also by the RENO experiment~\cite{Ahn:2012nd}
\begin{equation}
\sq = 0.113 \pm 0.013({\rm stat}) \pm 0.019 ({\rm syst}) \,.
\label{eq:RENO}
\end{equation}
These results confirm previous hints for a relatively large value of
$\theta_{13}$ from the T2K~\cite{Abe:2011sj} and Double
Chooz~\cite{Abe:2011fz} experiments, see also~\cite{Schwetz:2011zk}. Given
these exciting developments, the question arises whether there is a chance
to determine the neutrino mass ordering with currently running or planned
experiments. This possibility will have an important impact on the planning
and design of a subsequent generation of oscillation experiments towards the
search for leptonic CP violation.

The main goal of the current generation of accelerator experiments
(T2K~\cite{Itow:2001ee} and NOvA~\cite{Ayres:2004js}) as well as
reactor experiments (Double-Chooz \cite{Ardellier:2006mn}, RENO
\cite{RENO-cdr}, and Daya Bay \cite{Guo:2007ug}) is the determination
of $\theta_{13}$, see \cite{Mezzetto:2010zi} for a recent review. From
those experiments only NOvA may have sensitivity to the sign of
\dmq. However, the study performed in \cite{Huber:2009cw} shows that,
combining expected data from NOvA with the ones from all the other
mentioned experiments, only a poor sensitivity to the neutrino mass
ordering will be obtained: for $\sq \approx 0.09$ the sign of
\dmq\ can be determined at 90\%~CL in 2019 only for about 45\% of all
possible values of the CP phase $\delta$ and there is negligible
sensitivity at higher CL. Even for a fully optimised run time schedule
as well as optimistic upgrade assumptions for T2K and NOvA, the global
data from those experiments extrapolated until 2025 will provide
sensitivity at $3\sigma$ for only about 35\% of all $\delta$
values. For similar studies see also \cite{Huber:2002rs,
  Minakata:2003ca, Prakash:2012zx}.

Motivated by this situation we explore here the possibility to use data from
atmospheric neutrinos in order to determine the mass ordering. In
particular, in the India-based Neutrino Observatory (INO)~\cite{INO} a
magnetized iron calorimeter will be built for the observation of charge
separated muons induced from atmospheric neutrinos, with a 50~kt detector
expected to start data taking in 2017~\cite{INO-LP11}. A comparable small
sample of such atmospheric neutrino events has been observed by the MINOS
detector~\cite{Adamson:2005qc, Adamson:2007vt}. For large enough values of
$\theta_{13}$ matter effects will induce characteristic signatures for
atmospheric muon neutrinos as a function of the neutrino energy and zenith
angle, different for neutrinos and anti-neutrinos, depending on the mass
ordering, due to three-flavour matter effects \cite{Petcov:1998su,
Akhmedov:1998xq, Chizhov:1998ug, Chizhov:1999az}, see also
\cite{Akhmedov:2006hb}. The sensitivity of magnetized iron detectors to
those effects has been studied in~\cite{TabarellideFatis:2002ni,
Bernabeu:2003yp, PalomaresRuiz:2004tk, Indumathi:2004kd, Petcov:2005rv,
Samanta:2006sj, Kopp:2007ai, Gandhi:2007td, Samanta:2009qw}. Also non-magnetized
atmospheric neutrino detectors may provide sensitivity to the neutrino mass
ordering, see e.g., \cite{Gandhi:2007td, Kajita:2006gk, Fogli:2005cq, Huber:2005ep,
Campagne:2006yx, Gandhi:2008zs, Mena:2008rh}.  However, the
lack of event-by-event discrimination of neutrino and anti-neutrino induced
events leads to a dilution of the relevant signatures and therefore huge
detectors will be required \cite{Abe:2011ts, Autiero:2007zj}.

In the following we will investigate the impact of the INO atmospheric
neutrino data on the sensitivity to the neutrino mass ordering in a global
context by combining simulated data from INO, NOvA, and T2K. The remainder
of this work is organized as follows. In section~\ref{sec:details} we give
details of our simulation of atmospheric and accelerator data, as well as on
the statistical analysis. The main results are presented in
section~\ref{sec:results}, where we discuss the combined sensitivity of
atmospheric and accelerator data as a function of time for various
assumptions on the experimental configuration achieved in INO. In
section~\ref{sec:delta} we discuss the effect of neglecting the solar
mass-squared difference $\Delta m^2_{21}$ in the INO analysis. We conclude
in section~\ref{sec:conclusions}.

\section{Simulation details}
\label{sec:details}

\subsection{Atmospheric neutrinos in INO}

For the simulation of atmospheric data in INO we follow closely
\cite{Petcov:2005rv}, where technical details for the calculation of the
event rates, the used cross section and neutrino fluxes, as well as
the statistical analysis are given. Here we summarize our main
assumptions. We assume a muon threshold of 2~GeV and assume that muon
charge identification is perfect with an efficiency of 85\% above that
threshold. As stressed in \cite{Indumathi:2004kd, Petcov:2005rv} the
energy and direction reconstruction resolutions are crucial parameters
for the sensitivity to the mass ordering.  The ability to reconstruct
neutrino energy and direction depends on the corresponding resolutions
for the muon, the mean angle between muon and neutrino, and the energy
and momentum reconstruction for the hadronic shower. In the absence of
detailed Monte Carlo simulations we assume that neutrino energy and
direction resolutions are Gaussian with widths $\sigma_E$ and
$\sigma_\theta$, respectively, assuming two representative sets of
values, corresponding to a ``low'' or ``high'' resolution configuration: 
\be
\begin{split}
\sigma_E / E_\nu = 0.15 &\,,\qquad 
\sigma_\theta = 15^\circ
\qquad \text{(low)} \\
\sigma_E / E_\nu = 0.10 &\,,\qquad 
\sigma_\theta = 10^\circ
\qquad \text{(high)} 
\end{split}
\ee 
We take those resolutions to be independent of energy or zenith
angle. In a real experiment neither the resolutions nor charge ID
efficiencies will be constant, and one may expect different data
samples with diverse reconstruction qualities. Unfortunately detailed
reconstruction capabilities of the INO detector are currently not
available. Therefore we make the simplified assumptions stated
above.  Our results should be considered as a representative estimate
assuming that the adopted values can be achieved in average for the
majority of the events.  With a constant efficiency of 85\% and a
muon threshold of 2~GeV we find 242 (223) $\mu$-like events per 50~kt~yr exposure
for the high (low) resolution assuming no oscillations (sum of
neutrino and anti-neutrino events).

As shown in \cite{Petcov:2005rv} $e$-like events provide
additional sensitivity to the mass ordering. However, electrons and
positrons induce electromagnetic showers in an INO-like detector and the
reconstruction and charge separation of such events is difficult. Currently
it is not foreseen to consider $e$-like events in INO. Therefore we do not
include them in our analysis. 

We divide the simulated data into 20 bins in reconstructed neutrino
energy from 2~GeV to 10~GeV, as well as 20 bins in reconstructed
zenith angle from $\cos\theta = -1$ to $\cos\theta = -0.1$. We then
fit the two-dimensional event distribution in the $20\times 20$ bins
by using the appropriate $\chi^2$-definition for Poisson distributed
data, while also including the following systematic uncertainties: we assume a
20\% uncertainty on the over-all normalization of events, and 5\% on each of
the neutrino/anti-neutrino event ratio, the $\nu_\mu$ to $\nu_e$ flux ratio,
the zenith-angle dependence, and on the energy dependence of the
fluxes, see \cite{Petcov:2005rv} for details.

In the simulation of atmospheric neutrino data we set $\Delta m^2_{21}
= 0$. This is a reasonable approximation for neutrino energies above
2~GeV. We estimate the accuracy of this approximation in
section~\ref{sec:delta}. Under this assumption, atmospheric neutrino
data depend on the three parameters $\Delta m^2_\mathrm{eff},
\theta_{23}, \theta_{13}$. For the NOvA and T2K simulations we use
full three-flavour oscillation probabilities including $\Delta
m^2_{21}$ effects. In order to combine such simulations with INO we
use for the atmospheric analysis an effective mass-squared difference,
which is related to \dmq\ by \cite{Nunokawa:2005nx} (see also \cite{deGouvea:2005hk})
\be \label{eq:dmqeff}
\Delta m^2_\mathrm{eff} = \Delta m^2_{31} - (\cos^2\theta_{12} -
\cos\delta\sin\theta_{13}\sin 2 \theta_{12} \tan\theta_{23}) 
\Delta m^2_{21} \,.  
\ee 
This is particularly important when investigating the
sensitivity to the mass ordering, since to very good approximation the
degenerate solutions occur at $\pm \Delta m^2_\mathrm{eff}$, which
corresponds to slightly differing values of $|\dmq|$ according to the
above equation.

\subsection{NOvA, T2K, and reactor experiments}

For the simulation of the long-baseline accelerator experiments NOvA and T2K
we use the GLoBES software \cite{Huber:2004ka, Huber:2007ji} and follow
closely the analysis of \cite{Huber:2009cw}, where details on the assumed
experimental parameters can be found.  For T2K, we always assume the final
exposure corresponding to 5 years running with a beam power of 0.75~MW in
neutrinos only. Assuming a realistic beam power evolution, such an exposure
will be obtained around 2018.  For NOvA, the nominal exposure is 3~years for
neutrinos and anti-neutrinos each, with $6\cdot 10^{20}$~POT per year and a
15~kt detector. When considering the time evolution of the sensitivity we
assume the run plan for alternating neutrino and anti-neutrino beams as well
as the time evolution of the detector mass as shown in Fig.~11 of
\cite{Mezzetto:2010zi}. This should be considered as one particular
representative example, and in this scenario the nominal exposure
would be reached around 2019 for neutrinos and 2020 for anti-neutrinos. We
assume that the experiment is terminated after this date. 

The main sensitivity to the mass ordering comes from NOvA due to the
somewhat larger matter effect, because of the longer baseline of 810~km. The
matter effect is significantly smaller in T2K (295~km baseline) and
therefore T2K has no sensitivity to the mass ordering. However, T2K may
contribute indirectly by providing additional information on $\theta_{13}$
and the phase $\delta$. Therefore, we always consider the combination of
both beams in the following.

We include the determination of $\sq$ from reactor experiments as a
simple Gaussian prior centered at 0.09 with the current Daya Bay error
of 0.017 at 1$\sigma$ (systematic and statistical errors
combined). While improved data from the running reactor experiments
will tighten these bounds, the actual sensitivity of INO and NOvA to
the mass ordering does not depend crucially on the exact error.

\subsection{Parameter values and implementation}

\begin{figure}[t]
\centering \includegraphics[width=0.7\textwidth]{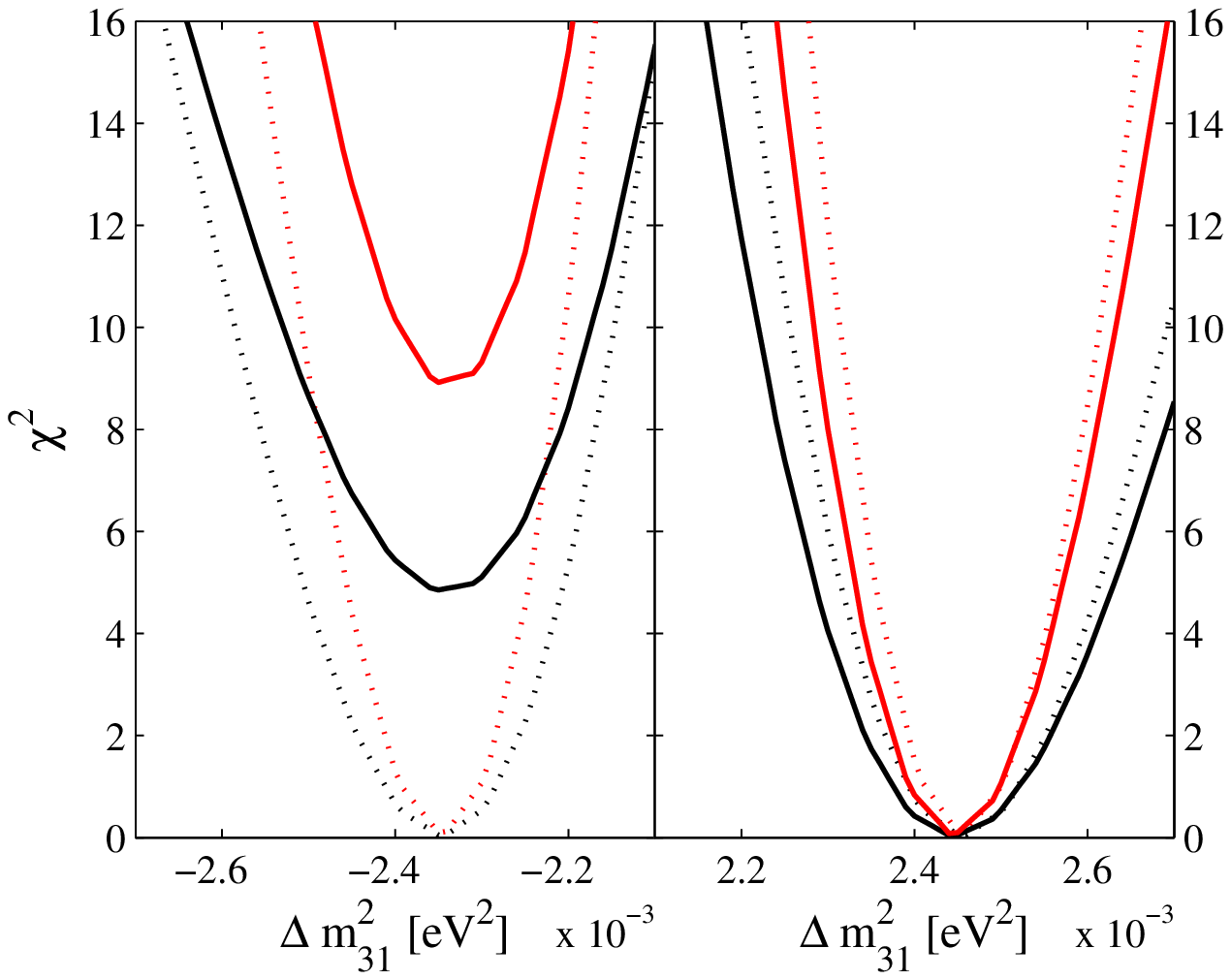}
\mycaption{The dependence of the marginalized atmospheric $\chi^2$ as
  a function of $\Delta m_{31}^2$ in the different mass orderings
  assuming $|\Delta m^2_{\rm eff}| = 2.4\cdot 10^{-3}$~eV$^2$. The dotted
  (solid) curves correspond to $\sq = 0$ (0.09) and the black (red)
  curves to a detector mass of 50~kt (100~kt). The running time has
  been assumed to be 10~years with the high resolution scenario.}
\label{fig:dm2chi2dep}
\end{figure}

For our simulations, we have fixed the values of the true neutrino oscillation parameters to
\begin{equation}
\begin{array}{l}
 \Delta m_{\rm eff}^2 = 2.4\cdot 10^{-3}~{\rm eV}^2, \
 \Delta m_{21}^2 = 7.8\cdot 10^{-5}~{\rm eV}^2, \\
 \theta_{23} = 45^\circ, \
 \theta_{12} = 33^\circ, \
 \sin^22\theta_{13} = 0.09,
\end{array}
\end{equation}
unless stated otherwise. When treating the accelerator experiments, the
sensitivity to the mass ordering depends significantly on the true values of
the CP phase $\delta$. We fix $\Delta m_{21}^2$ and $\theta_{12}$ in our
fits, since the impact of these parameters is expected to be minimal. Other
parameters are marginalized over, and we impose the following priors
(1$\sigma$ Gaussian errors): $\sigma(\sin^22\theta_{13}) = 0.017$ (the
uncertainty from Daya Bay~\cite{dayabay}), $\sigma(\Delta m^2_{31}) = 0.5
|\Delta m^2_{13}|$ (this is a very weak prior with the only purpose to guide
the minimization algorithm), and $\sigma(\theta_{23}) = 0.08\theta_{23}$. The
prior on $\theta_{23}$ corresponds to $\sigma(\sin^2\theta_{23}) \approx
0.063$ at $\theta_{23} = 45^\circ$, which is approximately the accuracy from
current data~\cite{Schwetz:2011qt}. Once T2K and NOvA are included in the
fit, they will provide a more accurate determination of $\theta_{23}$ than
this prior. By default we assume that the true mass ordering is normal and
test the sensitivity to exclude the inverted ordering. With the above
mentioned priors the results are very similar if the true ordering is
inverted, as we will show explicitly towards the end of
section~\ref{sec:results}. For each simulated value of the true parameters,
we find the minimum value of the $\chi^2$ in the opposite mass ordering.
When reporting sensitivity in terms of standard deviations we use one degree
of freedom to evaluate the $\chi^2$. Hence the number of $\sigma$ with which
the wrong mass ordering can be excluded is given by the square-root of the
$\Delta\chi^2$ between the two signs of $\dmq$.

In Fig.~\ref{fig:dm2chi2dep}, we show the behavior of the marginalized
$\chi^2$ as a function of $\Delta m_{31}^2$ for the case of atmospherics
only for $\sq = 0$ and 0.09, respectively.  As can be seen from the figure,
the mass orderings are indistinguishable when $\sq = 0$, while there is a
significant difference for $\sq = 0.09$. Also note the shifts in the best
fit of $\Delta m_{31}^2$ as compared to the simulated $\Delta m_{\rm
eff}^2$, which are well in agreement with Eq.~\ref{eq:dmqeff}.

Let us stress that the sensitivity of atmospheric data on the mass ordering
also depends on the value of $\theta_{23}$, with typically 
improved (weaker) sensitivity for $\theta_{23} > 45^\circ\, (< 45^\circ)$.
This effect has been investigated in \cite{Petcov:2005rv}. Here we always assume a
true value $\theta_{23} = 45^\circ$ in order represent the ``average''
sensitivity and briefly show this behavior in the end of the next section.

\section{Results}
\label{sec:results}

\begin{figure}[t]
\centering \includegraphics[width=0.7\textwidth]{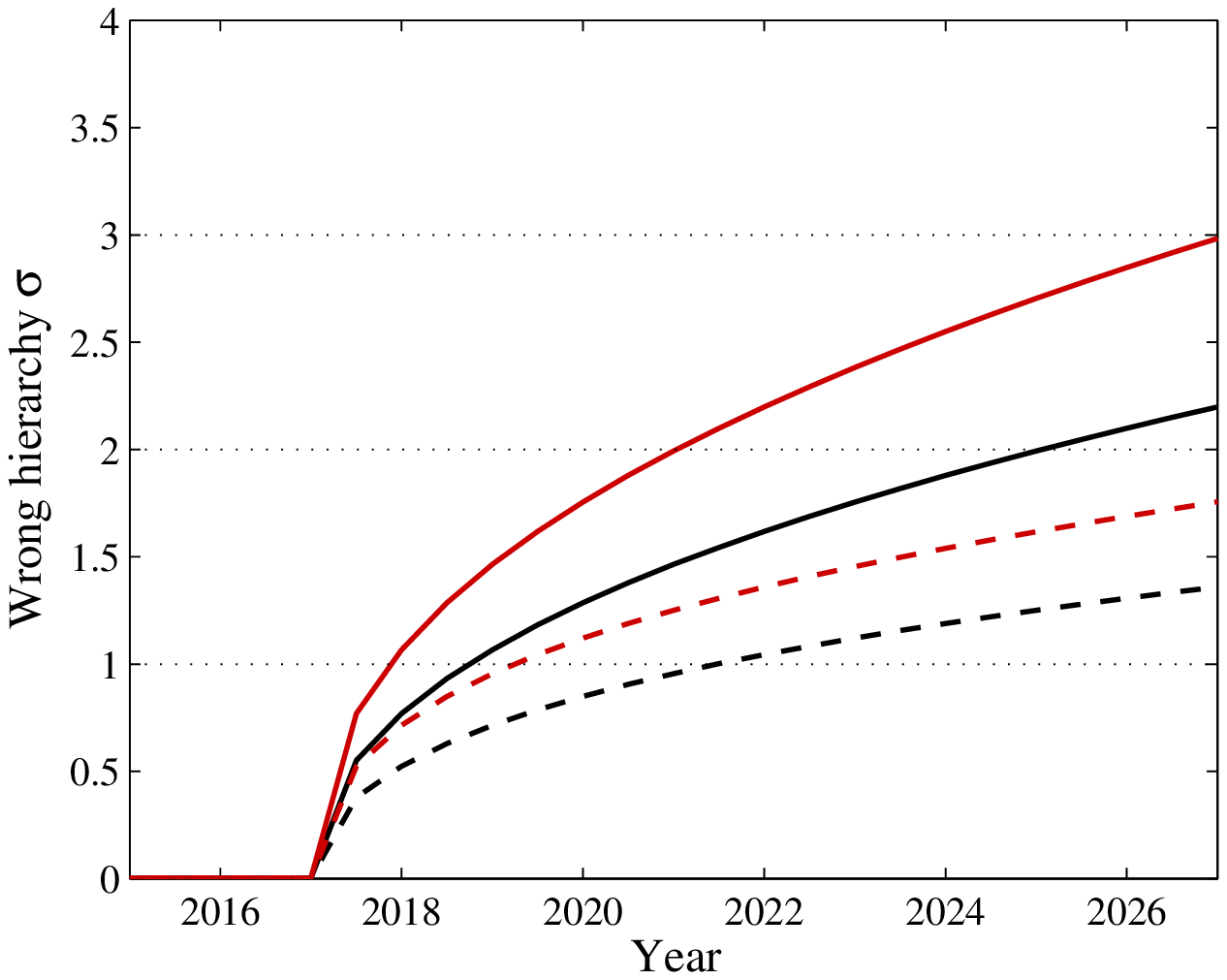} \mycaption{The
  sensitivity of an INO-like detector only as evolved in time assuming 
  start of data taking in the beginning of 2017 (ticks in the figure
  correspond to the beginning of the year). We show the number of standard
  deviations with which the wrong mass ordering can be excluded. Black (red)
  curves correspond to a detector mass of 50~kt (100~kt) and dashed (solid)
  curves correspond to the low (high) resolution scenario (see text). We
  assume true values $\sin^22\theta_{13}^\mathrm{tr} = 0.09$,
  $\theta_{23}^\mathrm{tr} = \pi/4$, and impose external priors at 1$\sigma$ of
  $\sigma(\sin^22\theta_{13}) = 0.017$ and $\sigma(\theta_{23}) =
  0.08\theta_{23}^\mathrm{tr}$.} \label{fig:TLatmo}
\end{figure}

Let us first investigate the sensitivity of atmospheric data alone.
In Fig.~\ref{fig:TLatmo}, we show how the sensitivity of INO only to
excluding the inverted ordering would evolve in time, depending on
implemented scenario. We would like to remind that to a
fairly good approximation this sensitivity is independent of $\delta$
and will be slightly worse if the effects of this parameter are
included (see Sec.~\ref{sec:delta}).
From this figure, we can deduce that the sensitivity of atmospheric
data would be strongly dependent on how well the atmospheric study can
be performed. In particular, in comparing the best and worse cases,
the high resolution 100~kt scenario would reach a 90\% CL sensitivity
after slightly less than two years of running, while the low
resolution 50~kt scenario would not accomplish this within the assumed
10~year lifetime of the experiment, after which the high resolution
100~kt scenario has reached a sensitivity close to $3\sigma$. Note
that the high resolution 50~kt scenario outperforms the low resolution
100~kt scenario, implying that reaching for an increase in resolution
may be preferable to increasing the detector mass
\cite{Indumathi:2004kd, Petcov:2005rv}, depending on the cost and
technical feasibility of doing so.

\begin{figure}[t]
\centering
  \includegraphics[width=0.48\textwidth]{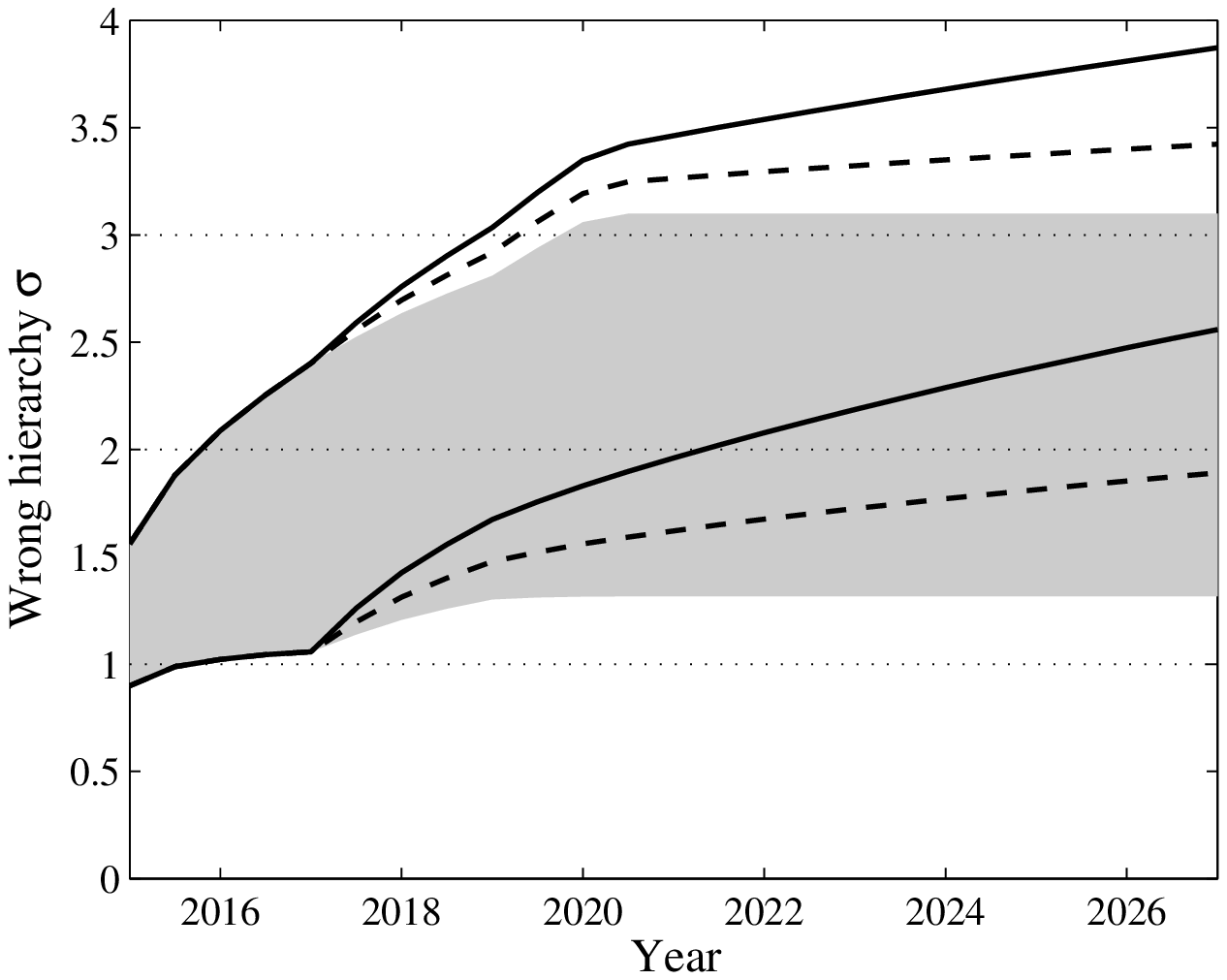}
  \includegraphics[width=0.48\textwidth]{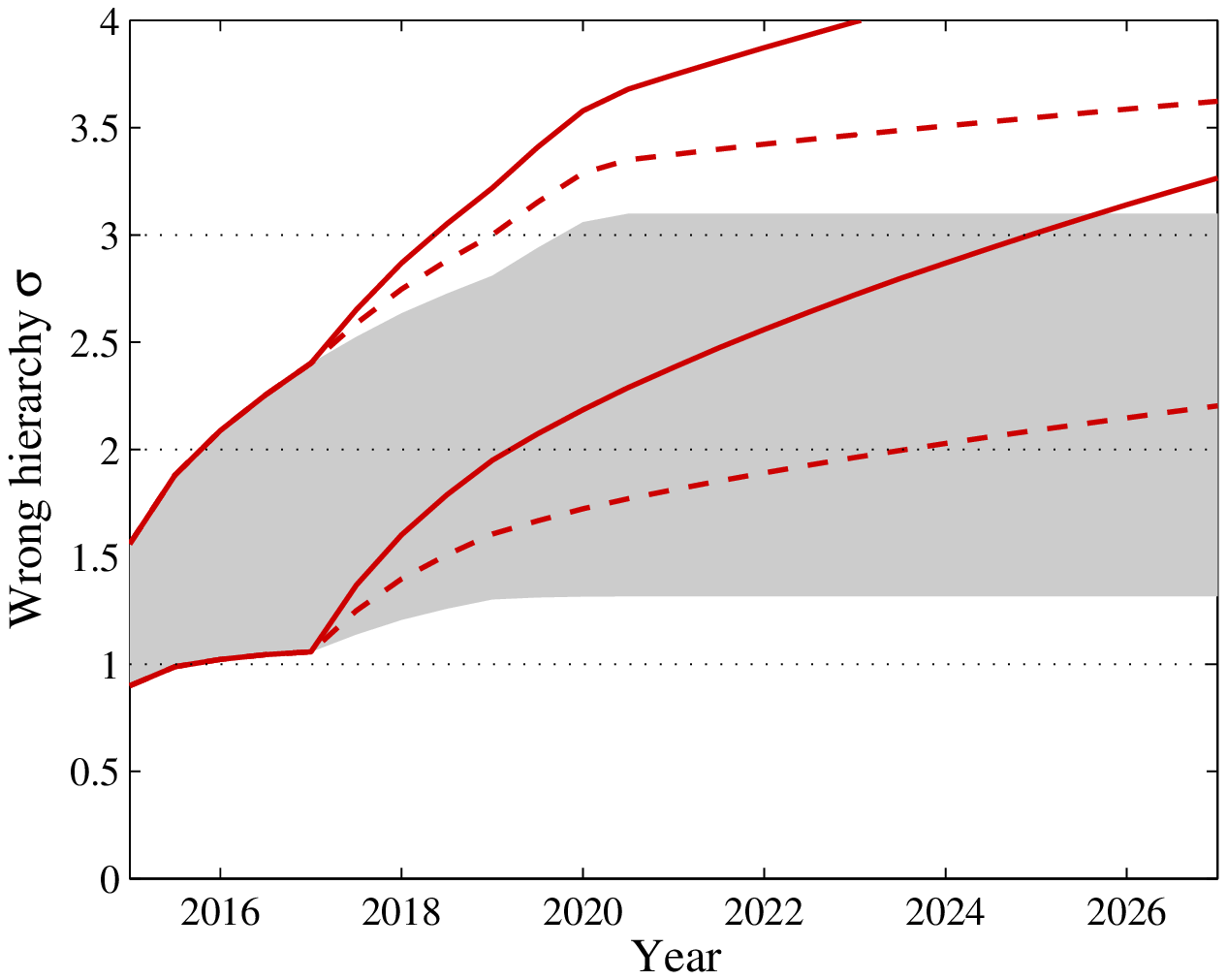}
  \mycaption{The minimum and maximum sensitivities (depending on the true
  value of $\delta$) of atmospherics combined with NOvA and T2K.  We show
  the number of standard deviations with which the wrong mass ordering can
  be excluded. The left (right) panel corresponds to a detector mass of
  50~kt (100~kt) and dashed (solid) curves correspond to the low (high)
  resolution scenario. The shaded area is the corresponding result for NOvA
  and T2K only. The true value of $\sq$ has been assumed to be $0.09$.}
  \label{fig:TLcomb}
\end{figure}

The sensitivity to exclude the inverted mass ordering by combining
information from INO together with NOvA and T2K is presented in
Fig.~\ref{fig:TLcomb}.  Since the sensitivity depends on the true value of
the CP violating phase $\delta$, we show both the maximum and minimum
sensitivities, corresponding to the most and least favourable values of
$\delta$, respectively. For comparison, we have also included the
sensitivity of the accelerator based experiments only. The kink in the least
favourable curve is due to the onset of anti-neutrino running in NOvA, which
in our example time line happens simultaneously with the onset of INO data
taking in 2017. From this figure, we see that in the low resolution
scenarios the sensitivity to the mass ordering is mildly improved compared
to that of the accelerator experiments only, increasing by roughly
$0.5\sigma - 1\sigma$ after the full 10~years of running, depending on the
scenario and the value of $\delta$. For these scenarios, the atmospheric
data only slightly adds information to the accelerator data during the
running time of NOvA, resulting in a given sensitivity to be reached
marginally ahead of when it would be reached by accelerator experiments
alone.

For the higher resolution scenarios, the rise in the precision is
significantly faster, in particular during the time when atmospheric
and accelerator experiments are running in parallel. This results in a
significant improvement in the sensitivities in a relatively short
time span. It should be noted that this effect is present both for the
values of $\delta$ to which accelerator experiments are most
sensitive, as well as those values to which they are least sensitive,
leading to an overall increase in the sensitivity irrespective of the
value of $\delta$. The statistical level where the synergy between
atmospheric and accelerator experiments is most apparent is around the
$2\sigma$ region for the least favourable values of $\delta$, where
the final sensitivity is impacted by both types of experiment. Most
noticeably, NOvA and T2K never reach this sensitivity by themselves,
while the high resolution atmospheric data by itself would reach it
several years later than the combined. Most strikingly, the combined
sensitivity in the most optimistic high resolution 100~kt scenario
would reach the $2\sigma$ within 2~years of the start of atmospheric
data taking, still within the lifetime of NOvA.

\begin{figure}[t]
\centering \includegraphics[width=0.7\textwidth]{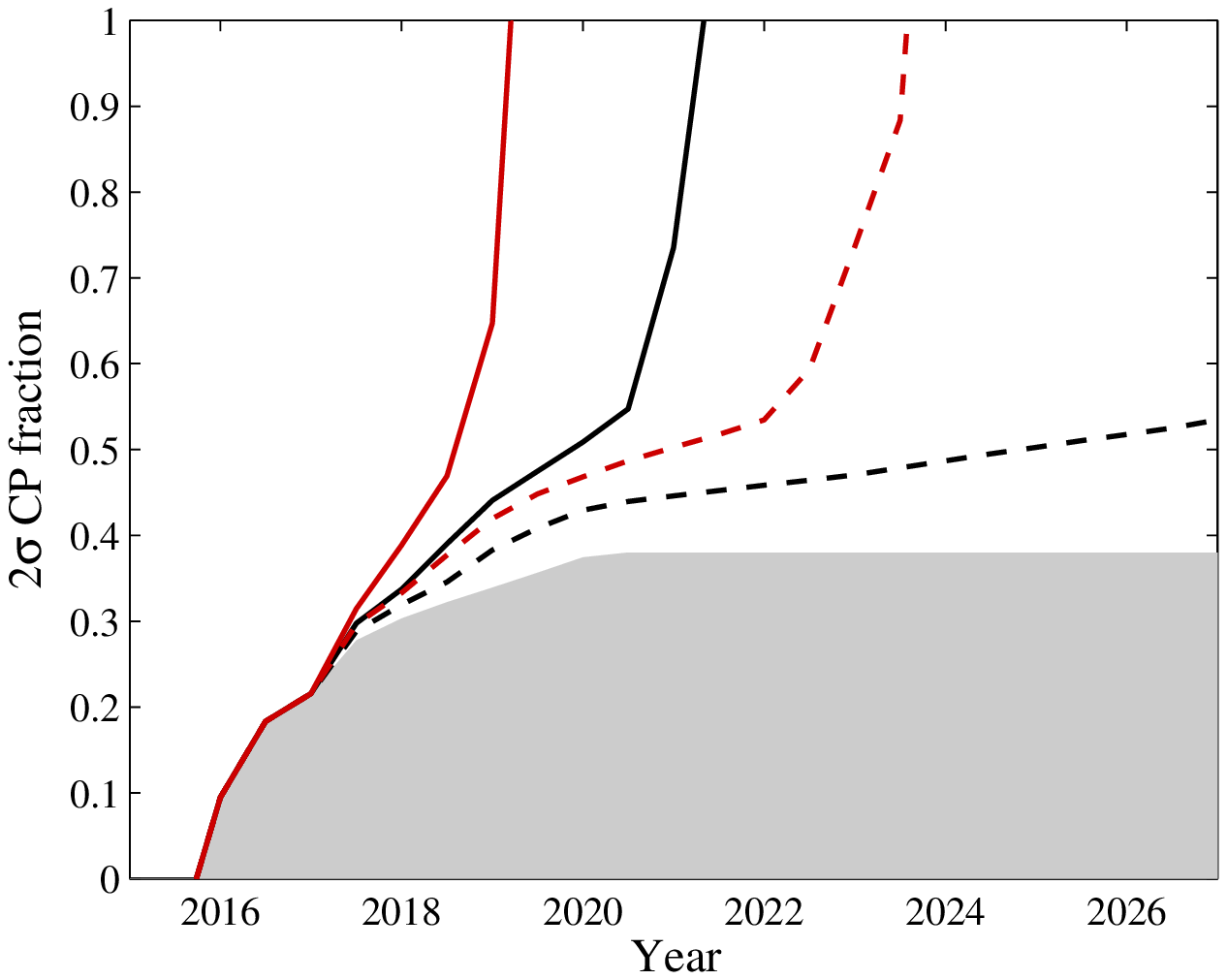}
\mycaption{The time evolution of the fraction of values of the CP
  violating phase $\delta$ for which the combination of INO, NOvA, and T2K
  would be sensitive to the mass ordering at $2\sigma$.
  Black (red) curves correspond to a detector mass of 50~kt (100~kt)
  and dashed (solid) curves correspond to the low (high) resolution
  scenario. The shaded area is the corresponding result for NOvA and T2K only.
  The true value of $\sq$ has been assumed to be $0.09$.}
\label{fig:TLcpfrac}
\end{figure}

\begin{figure}[!h]
\centering 
\includegraphics[width=0.48\textwidth]{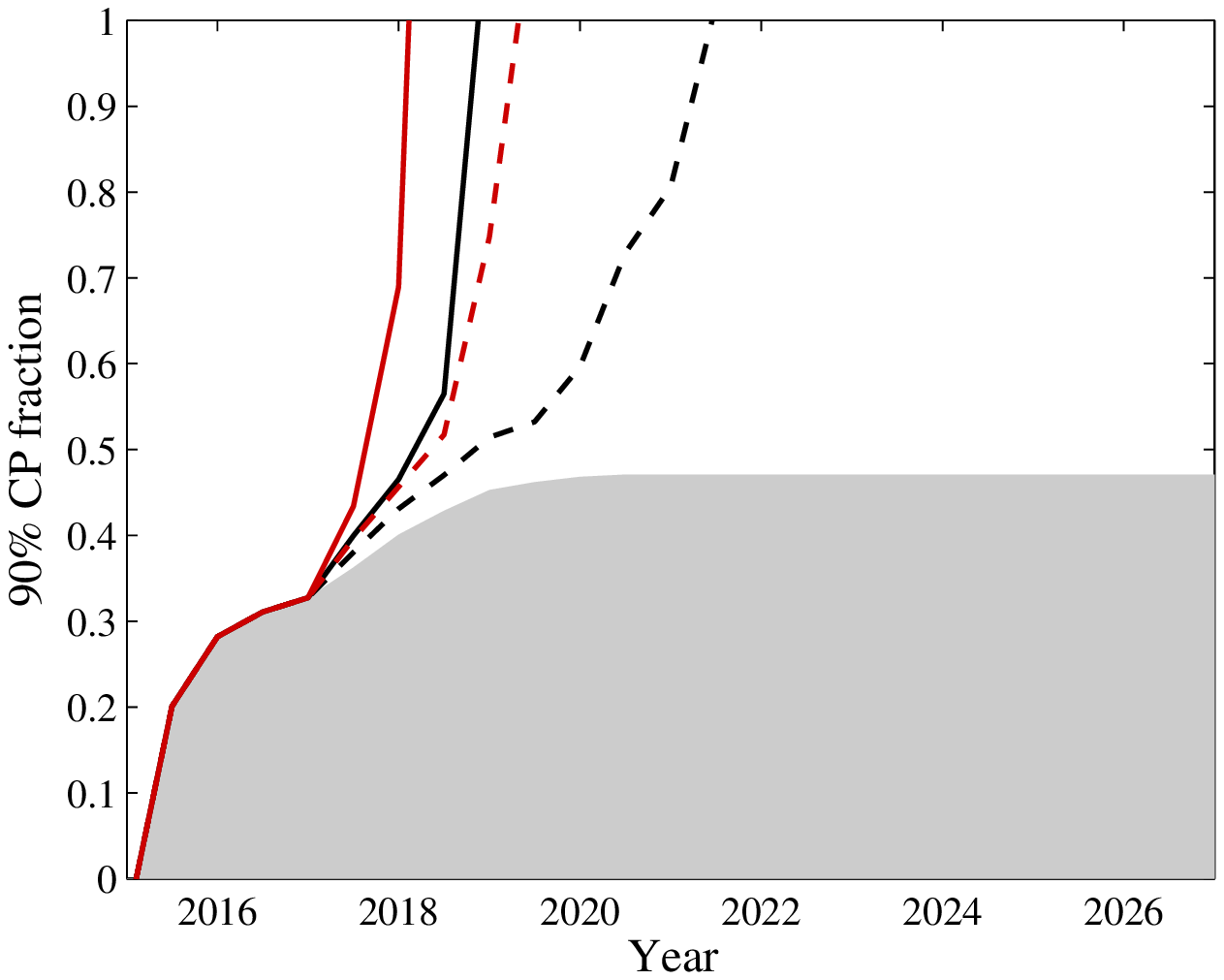} 
\includegraphics[width=0.48\textwidth]{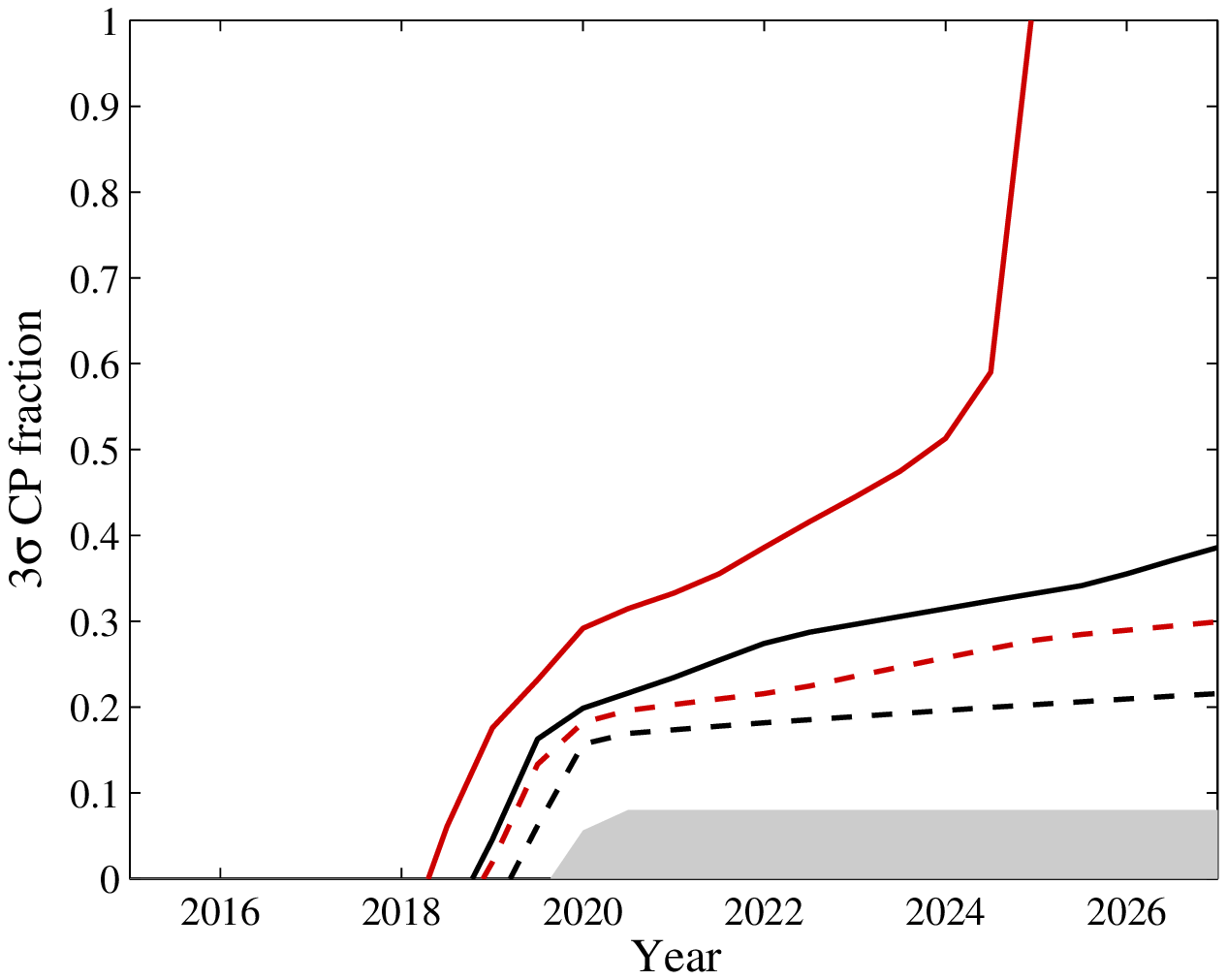}
\mycaption{Same as Fig.~\ref{fig:TLcpfrac} but for sensitivity at the 90\% ($3\sigma$) CL for the left (right) panel.}
\label{fig:TLcpfracOther}
\end{figure}

In Fig.~\ref{fig:TLcpfrac}, we show the time evolution of the
$2\sigma$ CP fraction, the fraction of the values of true $\delta$
(assuming a flat distribution) for which the sensitivity to the mass
ordering is $2\sigma$ or better. Note that this information is
complementary to that presented in the previous figure, as it also
contains information about the sensitivity dependence on $\delta$, not
just the maximum and minimum sensitivity.  Comparing these two
figures, we can deduce that unfortunately there are more values of
$\delta$ which are close to the minimal sensitivity than to the
maximal one. In particular, for the worst case of low resolution and a
50~kt detector, the $2\sigma$ CP fraction remains around 0.5 by 2027,
even though the minimal sensitivity at this point is around
$1.9\sigma$. This is further illustrated by the rate at which the CP
fraction grows after reaching 0.5 for the other scenarios. For
completeness, we also show the 90\% and $3\sigma$ CP fractions in
Fig.~\ref{fig:TLcpfracOther}. We deduce from these figures that it
would be relatively easy to obtain a 90\%~CL hint for rejecting the
inverted ordering, while a $3\sigma$ evidence will be significantly
more challenging. However, in both cases, the interplay between
atmospheric and accelerator data will allow for establishing these at
an earlier time, regardless of the true value of $\delta$ or the
confidence level in question.

\begin{figure}[t]
\centering
  \includegraphics[width=0.7\textwidth]{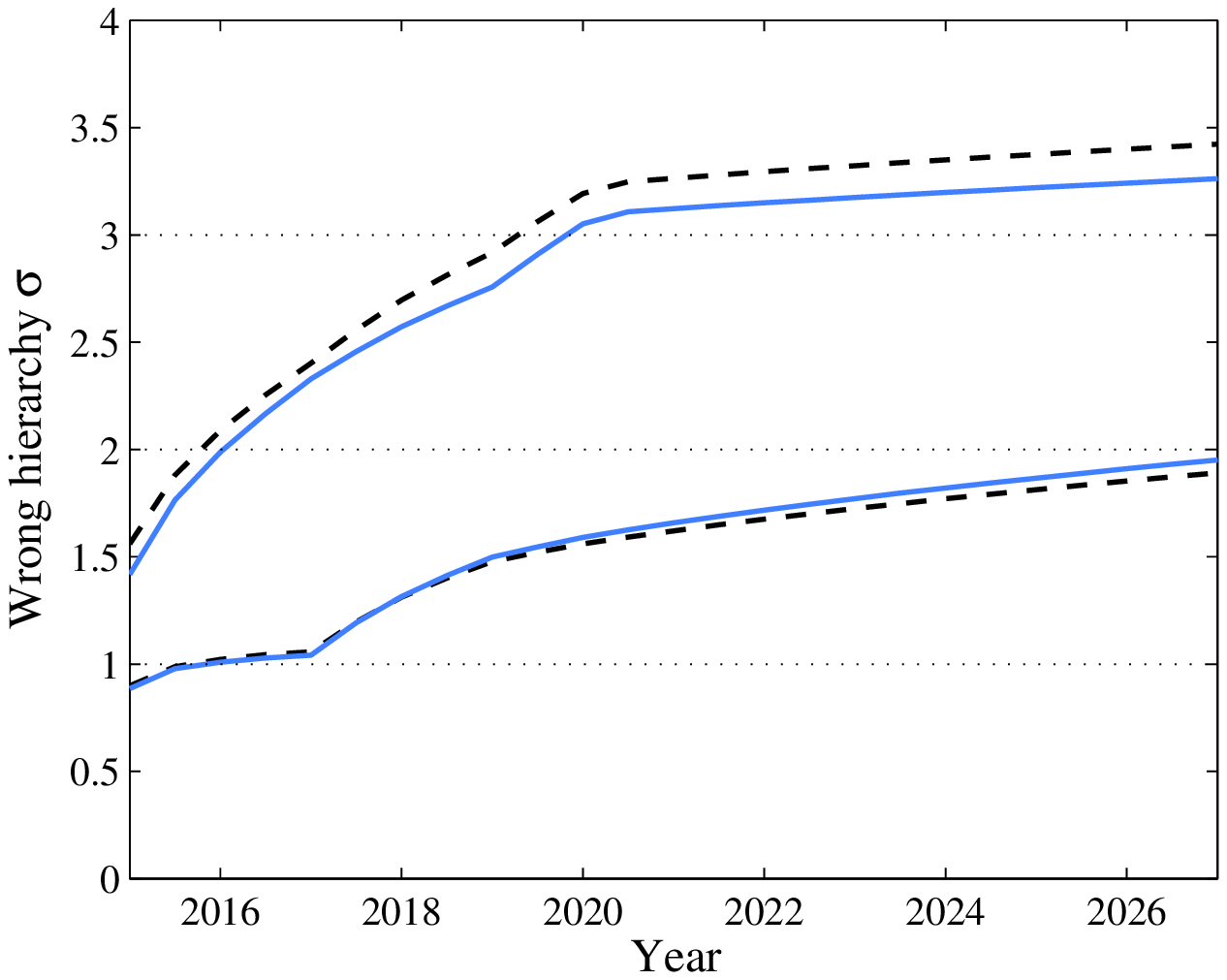}
  \mycaption{The minimum and maximum sensitivities (depending on the true
  value of $\delta$) of atmospherics combined with NOvA and T2K assuming
  that the true mass ordering is normal (inverted) for dashed-black (solid-blue).
  We assume a detector mass of 50~kt, low resolution, and a true value 
  $\sq = 0.09$.}
  \label{fig:norm-inv}
\end{figure}

So far we always assumed that the true mass ordering is normal. In
Fig.~\ref{fig:norm-inv} we compare the difference in sensitivity between a
true normal or inverted mass ordering, confirming that the sensitivity is
very similar. The features induced by the matter resonance will appear in
the data either for neutrinos or anti-neutrinos, depending on whether the
true ordering is normal or inverted, respectively. While statistics will be
larger for neutrinos than for anti-neutrinos, the important quantity for the
sensitivity is the {\it difference} in event numbers expected for the
different mass orderings in each of the two samples, and this difference is
largely independent of the true ordering. This statement is true for
approximately fixed oscillation parameters, especially $\theta_{13}$. Our
simulation shows that the accuracy on $\theta_{13}$ and other parameters as
provided by current Daya Bay data as well as the simulated data from T2K and
NOvA, is sufficient for the above argument to be valid, in agreement with
\cite{Petcov:2005rv}. For Fig.~\ref{fig:norm-inv} we have assumed a 50~kt
detector mass and low resolution for INO. Corresponding results for all
other configurations considered in this work are very similar to the one
shown in Fig.~\ref{fig:norm-inv}.

\begin{figure}
\centering 
  \includegraphics[width=0.48\textwidth]{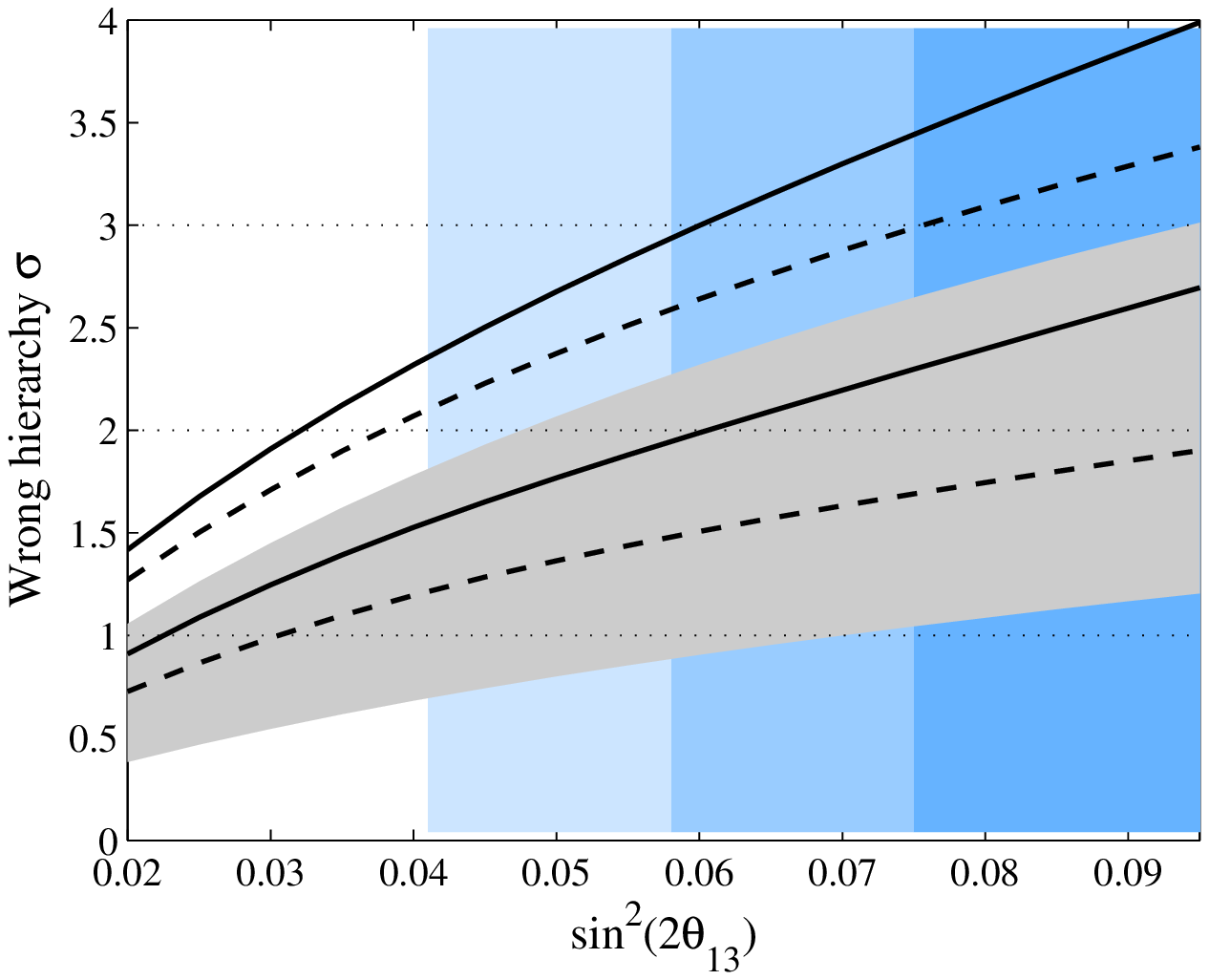}
  \includegraphics[width=0.48\textwidth]{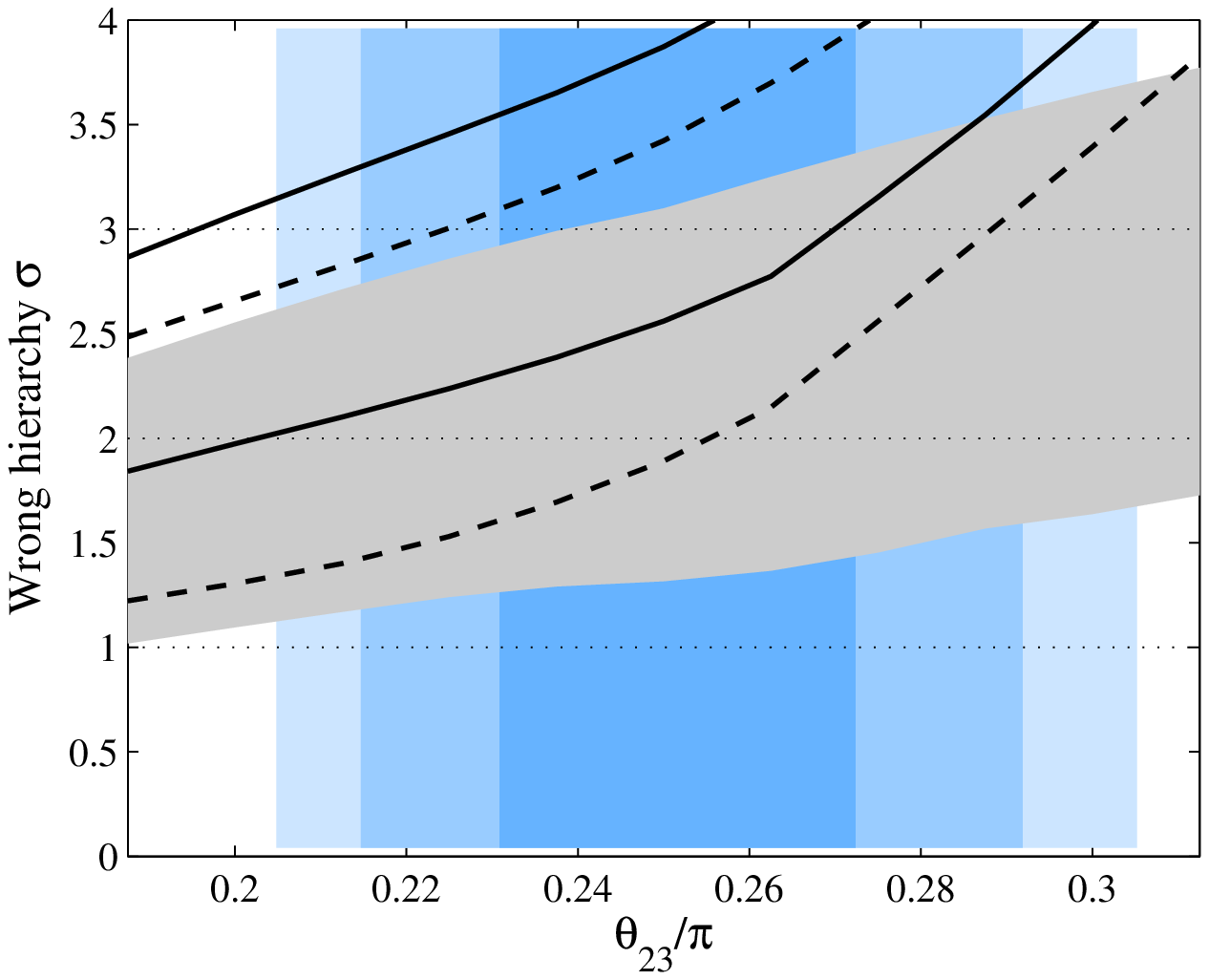} 
  \mycaption{The minimum and maximum sensitivities (depending on the true
  value of $\delta$) of atmospherics combined with NOvA and T2K as a function of the true value 
  of $\theta_{13}$ (left) and $\theta_{23}$ (right), respectively.  We show
  the number of standard deviations with which the wrong mass ordering can
  be excluded. We have assumed the
  total final exposures for NOvA and T2K, and an exposure of 500~kt~yr for
  INO. Dashed (solid) curves correspond to the
  low (high) resolution scenario. The gray-shaded area is the corresponding
  result for NOvA and T2K only. The blue-shaded areas indicate the current
  1, 2, 3$\sigma$ regions of the parameters (from dark to light shading).} \label{fig:th13}
\end{figure}

Despite the strong evidence for a non-zero (and relatively large) value for
$\theta_{13}$ the current uncertainty from Daya Bay~\cite{dayabay} still allows
for a significant spread in $\sq$, see Eq.~\ref{eq:th13}. In
Fig.~\ref{fig:th13} we show the dependence of the sensitivity to the mass
ordering as a function of the true value of $\sq$, assuming the final
exposure for all experiments. The blue shading indicates the 1, 2, and
3$\sigma$ lower bounds on $\sq$ according to Eq.~\ref{eq:th13}. We observe
that if the true value of $\theta_{13}$ happens to be close to the current
$3\sigma$ lower bound, the sensitivity to the mass ordering will be
noticeably weaker than at the current best fit point, typically allowing for
an exclusion of the wrong hierarchy with about one standard deviation
smaller significance. Correspondingly better sensitivities can be achieved
for $\theta_{13}$ larger than the current best fit point (not shown).
In a similar fashion, we also show the dependence of the sensitivity on the
mixing angle $\theta_{23}$. As shown in \cite{Petcov:2005rv}, the
sensitivity becomes better for values of $\theta_{23}$ above $\pi/4$ and
worse for smaller values, see e.g., Eq.~32 of \cite{Petcov:2005rv}. It is
also notable that for smaller values the atmospherics help only marginally,
while for larger values the additional information from atmospherics have a
larger impact.

\section{On the size of $\Delta m^2_{21}$ effects in INO}
\label{sec:delta}

For our simulations of INO data we neglect effects of $\Delta m^2_{21}$.
Oscillations of atmospheric neutrinos due to the solar mass-squared
difference are important for neutrino energies below 1~GeV, and since we
always impose a threshold of 2~GeV $\Delta m^2_{21}$ effects are expected to
be small. The approximation $\Delta m^2_{21} =0$ greatly simplifies the
numerical calculations and makes the detailed analysis including parameter
correlations, detector resolutions, as well as systematical uncertainties
feasible. In this section we check the accuracy of this approximation by
considering a simplified analysis taking into account full three-flavour
oscillation probabilities.

\begin{figure}[t]
\centering \includegraphics[width=0.5\textwidth]{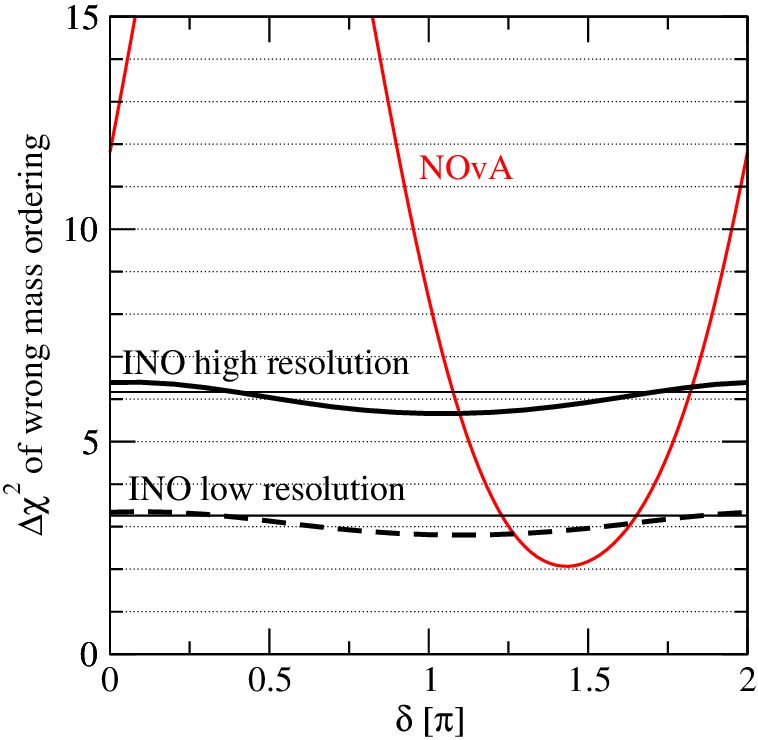}
\mycaption{$\Delta\chi^2$ of the wrong mass ordering for $\sq = 0.09$
  including a finite $\Delta m^2_{21}$ in the INO simulation as a
  function of the CP phase $\delta$. Data is
  simulated for normal mass ordering and $\delta = 0$. Those data are
  fitted assuming the inverted mass ordering varying $\delta$. When
  changing the sign of $\dmq$ we keep $|\Delta m^2_\mathrm{eff}|$
  constant, see Eq.~\ref{eq:dmqeff}, and all other oscillation
  parameters are fixed. For INO only statistical errors are taken into
  account, corresponding to an exposure of 500~kt~yr, and we show
  results assuming high and low resolutions. The horizontal thin lines
  correspond to the $\Delta\chi^2$ for $\Delta m^2_{21} = 0$. For NOvA
  we assume 3~yr nominal exposure for neutrinos as well as
  anti-neutrinos.}
\label{fig:delta}
\end{figure}

We consider only statistical errors corresponding to a 500~kt~yr INO
exposure and neglect systematical uncertainties in the fit. We
simulate ``data'' assuming normal mass ordering, $\sq = 0.09$, and
$\delta = 0$. Those ``data'' are fitted with inverted mass ordering,
where we keep $|\Delta m^2_\mathrm{eff}|$ constant according to
Eq.~\ref{eq:dmqeff}. All other oscillation parameters are fixed at the
values assumed in generating the data, except for the CP phase
$\delta$. The $\Delta \chi^2$ is shown in fig.~\ref{fig:delta} for low
and high resolutions in INO. We see that $\chi^2$ changes by about one
unit as a function of $\delta$. In the figure we also show the $\Delta
\chi^2$ value obtained by assuming $\Delta m^2_{21} = 0$. We conclude
that full three-flavour effects do not lead to additional sensitivity
to the mass ordering, but will diminish the sensitivity 
by less than one unit in $\chi^2$ when marginalized over $\delta$. This is true
for both assumptions on the resolution. In this sense our INO
sensitivities are slightly optimistic. One can expect that those
results hold also in the presence of systematical effects.

The various assumptions we had to make at the current stage on
reconstruction capabilities and efficiencies introduce uncertainties on the
sensitivity which are most likely larger than the one from neglecting the
solar mass-splitting, justifying our approximation. However, we stress that
once reliable detector properties become available it will be necessary to
include full three flavour effects in order to obtain accurate sensitivity
predictions.

In Fig.~\ref{fig:delta} we show also the $\delta$ dependence of the
$\chi^2$ from NOvA data for the same type of analysis. We see that for
NOvA three-flavour effects are essential and there is a significant
dependence on the CP phase $\delta$.\footnote{The NOvA curve in
  Fig.~\ref{fig:delta} depends strongly on the true value of $\delta$,
  whereas the result for INO depends only weakly on the assumed value.}
Comparing the variation of INO and NOvA we conclude that there is
very little complementarity with respect to $\delta$ dependent
effects. The $\delta$ best fit point will be completely dominated by NOvA, and
hence the small decrease in sensitivity for INO might be somewhat
lifted once combined with NOvA. This results in $\chi^2$ values very similar
to the one obtained in the approximation $\Delta m^2_{21} = 0$.

\section{Conclusions}
\label{sec:conclusions}

The recent discovery of a relatively large value of $\theta_{13}$ by
the Daya Bay reactor experiment opens exciting possibilities for the
future neutrino oscillation program. In this paper we have focused on
the determination of the neutrino mass ordering being normal, $\dmq >
0$, or inverted, $\dmq < 0$. Currently planned long-baseline
accelerator experiments, in particular NOvA and T2K, only have a poor
sensitivity to the mass ordering even in case of large
$\theta_{13}$. On the other hand, a large $\theta_{13}$ provides
potentially interesting opportunities for the INO atmospheric neutrino
experiment, which is planning to start data taking in 2017. We have
investigated the combined sensitivity of these experiments for different
assumptions on the size and event reconstruction capabilities of INO.

Assuming ``low'' resolutions (15\% and $15^\circ$ reconstruction accuracy in
neutrino energy and direction, respectively) and a 50~kt detector (fiducial)
only a poor global sensitivity is obtained and there is only marginal
improvement from combining INO with NOvA and T2K data. Significant synergy
and improved sensitivity is obtained for ``high'' resolution (10\% and
$10^\circ$) and/or doubling the INO detector mass. We find that improving
the resolution is more effective than increasing the detector mass. For high
resolution we find that a $2\sigma$ determination of the mass ordering is
possible irrespective of the CP phase $\delta$ in 2021 (2019) with a 50~kt
(100~kt) detector. The high resolution 100~kt detector even allows a
determination at $3\sigma$ around 2025. These conclusions hold for $\sq =
0.09$, close to the current best fit point. The sensitivity of the
considered experiments still depends crucially on the actual true value of
$\theta_{13}$ within the currently allowed $3\sigma$ region, and to a lesser
extent also on $\theta_{23}$.

In our analysis of simulated INO data we assumed a constant efficiency of
85\%, and energy and direction resolutions to be independent of energy and
direction. These are simplifying assumptions and once a detailed detector
simulation becomes available a more realistic analysis should be carried
out. We have also discussed the impact of effects related to the solar-mass
splitting $\Delta m^2_{21}$ and the CP phase $\delta$ for the atmospheric
neutrino data and have shown that their impact on the sensitivity to the
mass ordering is small.

In conclusion, the by now established large value of $\theta_{13}$ opens the
possibility to determine the neutrino mass ordering within a time frame of
about ten years with experiments currently under construction. Atmospheric
neutrino data from INO may be crucial in order to achieve this goal and we
believe that it is important to include such synergies in the global context
towards future neutrino oscillation facilities. Our study suggest, however,
that in order to achieve a relevant sensitivity some improvements of the INO
detector (either in event reconstruction capabilities, detector mass, or both)
seem to be necessary. Complementary information could be provided by $e$-like
events, and the sensitivity of INO could be potentially increased
significantly if the reconstruction of charge-separated $e$-like events was
possible \cite{Petcov:2005rv}. 

\bigskip
{\bf Acknowledgement.} We thank Tarak Thakore, Nita Sinha, and Enrique
Fernandez-Martinez for useful discussions and Anselmo Meregaglia
for spotting a typo in Eq.~\ref{eq:dmqeff} in earlier versions of this
paper.

\end{document}